\documentstyle[12pt,aaspp4]{article} 
\def\tempest%
{\begin{array}{ccc} 
1 & 1 & 1 \\ 
1 & 1 & 1 \\ 
4 & 3 & 8 
\end{array}}

\def\kms{{\rm km}\,{\rm s}^{-1}} 
 
\def\gev{{\rm GeV}}

\begin{document}

\title{Can Heavy WIMPs Be Captured by the Earth?}
\author 
{Andrew Gould and S.M.\ Khairul Alam}
\affil{Ohio State University, Department of Astronomy, Columbus, OH 43210} 
\affil{E-mail: gould, alam@astronomy.ohio-state.edu} 
\begin{abstract} 

	If weakly interacting massive particles (WIMPs) in bound solar
orbits are systematically driven into the Sun by solar-system resonances
(as Farinella et al.\ have
shown is the case for many Earth-crossing asteroids), then the capture
of high-mass WIMPs by the Earth would be affected dramatically because
high-mass WIMPs are captured primarily from bound orbits.  WIMP
capture would be eliminated for $M_x>630\,\gev$ and would be highly
suppressed for $M_x\ga 150\,\gev$.  Annihilation of captured WIMPs and
anti-WIMPs is expected to give rise to neutrinos coming from the Earth's
center.  The absence of such a neutrino signal has been used to place limits
on WIMP parameters.  At present, one does not know if typical WIMP
orbits are in fact affected by these resonances.  Until this question is
investigated and resolved, 
one must (conservatively) assume that they are.  Hence, limits
on high-mass WIMP parameters are significantly weaker than previously 
believed.

\keywords{asteroids -- elementary particles -- dark matter -- Galaxy: halo}
\end{abstract}

\section{Introduction} 

If weakly interacting massive particles (WIMPs) comprise the dark matter, 
they would have a local density $\rho_x \sim 0.3\,\gev\,\rm cm^{-3}$.
They would be potentially detectable in several ways including directly
in underground detectors, from WIMP-anti-WIMP annihilations in the Galactic
halo, and inidrectly from neutrinos produced by annihilations of WIMPs
captured by the Earth and the Sun (e.g., 
Jungman, Kamionkowski, \& Griest 1996).

	WIMP capture by the Sun was first discussed in the context of a
proposed WIMP solution to the solar-neutrino problem (Press \& Spergel 1985; 
Faulkner \& Gilliland 1985).  Subsequently, several workers realized that if
both WIMPs and anti-WIMPs were captured by the Sun (Silk, Olive, \& Srednicki
1985; Gaisser, Steigman, \& Tilav 1986; Srednicki, Olive, \& Silk 1987;
Griest \& Seckel 1987) or Earth (Freese 1986; Krauss, Srednicki, \& Wilczek
1986; Gaisser et al.\ 1986), they might annihilate into neutrinos which would
then pass directly through these boides and into neutrino detectors near
the surface of the Earth.  The failure to detect neutrinos coming from
the centers of the Sun and Earth could then place constraints on the
WIMP mass $M_x$ and cross section $\sigma_x$.

	Because the Sun's gravitational potential is much deeper than the
Earth's, it initially appeared as though the Sun would be so much more
effective at capture that the annihilation signal from the Sun would always 
be much larger than from the Earth.  However, Gould (1987) showed that there
were ``resonant enhancements'' in capture by the Earth whenever the WIMP
mass $M_x$ was close to the mass $M_A$ of an element $A$ that is common in
the Earth.  Specifically, for hard-sphere cross sections $\sigma_{xA}$, the
capture rate is given by (see also Gould 1992)
\begin{equation}
C ={N_A \sigma_{xA}\over \beta_-}\int_0^{v_{\rm cut}} 
d^3 u (v_{\rm cut}^2 - u^2){f({\bf u})\over u}
\label{eqn:capture1}   
\end{equation}
where 
\begin{equation}
v_{\rm cut}^2\equiv \beta_- v_{\rm esc}^2,\qquad \beta_\pm\equiv
{4 M_x M_A\over (M_x \pm M_A)^2},
 \label{eqn:vcutdef}   
\end{equation}
$f({\bf u})$ is the distribution of velocities $\bf u$ of WIMPs relative to
the Earth in the terrestrial neighborhood but away from the Earth's potential, 
$N_A$ is the total number of atoms
$A$ in the Earth, and $v_{\rm esc}\sim 13\,\kms$ is the escape velocity at the
position of the atoms.  In principle, one should allow for a range of
escape velocities of the atoms, but for the case of iron in the Earth
(of interest here), Gould (1987) showed that the error induced by simply
using the average escape velocity is small.

	In the present context, it is very important to note that $v_{\rm cut}$
is the maximum velocity for which an ambient WIMP can be captured by
the Earth.  If $M_A\sim M_x$, then $\beta_-\gg 1$ and so 
$v_{\rm cut}\gg v_{\rm esc}$.  In this case, capture is dominated by WIMPs
that have velocities typical of the halo $v_h\sim 300\,\kms$.  This is the
source of the ``resonant enhancements'' mentioned above: over the whole
mass range $12\,\gev\la M_x \la 65\,\gev$, WIMP capture is dominated by
resonances with elements common in the Earth, oxygen, magnesium, silicon,
and iron.  See Figure 2 from Gould (1987).  
For masses $M_x\ga 65\,\gev$, WIMP capture is overwhelmingly due to the
``tail'' of the iron resonance.  If the velocity distribution at
low velocities could be approximated as a constant $f({\bf u})\simeq f(0)$,
then the capture rate would take the form
\begin{equation}
C_0 = {\pi N_A \sigma_{xA}f(0)v_{\rm cut}^4\over \beta_-}
\rightarrow 4 \pi N_A \sigma_{xA}f(0)v_{\rm esc}^4\,{M_A\over M_x}
\qquad {\rm [assuming}\ f({\bf u})=f(0)\ {\rm for}\ u<v_{\rm cut}].
 \label{eqn:cap0}   
\end{equation}
Thus in the high-mass limit (evaluated after the arrow), the capture rate
would fall $C_0\propto M_x^{-2}$: one power of $M_x$ appears explicitly and
the other is implicit in $f({\bf u})$ for fixed $\rho_x$.

	Gould (1988) showed that, unfortunately, the high-mass capture rate
might not be so simple.  If one restricts attention to WIMPs that are
{\it not} bound tothe Sun, then
\begin{equation}
f_{\rm unbound}({\bf u}) = 0 \qquad {\rm for}\ |{\bf u} + {\bf v}_\oplus|
\leq 2^{1/2}v_\oplus,
\label{eqn:hole1}   
\end{equation}
where ${\bf v}_\oplus$ is the velocity of the Earth.  To properly
evaluate capture in the high-mass regime it is therefore necessary to evaluate 
the velocity distribution of bound as well as unbound WIMPs.  In particular, if
there were no bound WIMPs, then there would be no WIMP capture at all
for WIMPs with $v_{\rm cut}\leq (2^{1/2}-1)v_\oplus$, and WIMP capture
would be highly suppressed for WIMPs with $v_{\rm cut}\la v_\oplus$.  These
thresholds correspond to $M_x\sim 320\,\gev$ and $M_x\sim 120\,\gev$ 
respectively.

	Gould (1991) argued from detailed balance that regardless of the 
initial WIMP velocity distribution at the time of the formation of the solar
system, $f_{\rm bound}({\bf u})$ would be driven toward $f_\odot(0)$, the
low-velocity limit of the velocity distribution in the solar neighborhood
but away from the solar potential,
\begin{equation}
f_{\rm bound}({\bf u})\rightarrow f_\odot(0)\qquad {\rm (Gould}\ 1991).  
 \label{eqn:gould1991}   
\end{equation}
He found that the charactersitic time for the evolution of the velocity 
distribution is less than the age of the solar system for $u\la v_\oplus$
(Fig.\ 3 from Gould 1991).  While these arguments still left indeterminate
the bound WIMP distribution for $u\ga v_\oplus$, in practice the capture
rate would not be seriously affected for any mass even if all these
orbits were empty.  Hence Gould's (1991) arguments appeared to resolve the
problem of WIMP capture for any $M_x$, $\sigma_{xA}$ and Galactic WIMP
distribution.

\section{New Developments}

	There have been two new developments since 1991 that challenge this
seemingly closed case.  First, Farinella et al.\ (1994) have shown by
direct numerical integration that a large fraction of Earth-crossing asteroids
are systematically driven into the Sun by various solar-system resonances.
Gladman et al.\ (1997) and Migliorini et al.\ (1998) 
have further studied this
problem and generally confirmed the initial results.  These solar collisions
occur on Myr time scales, several orders of magnitude faster than the 
characteristic diffusion times evaluated by Gould (1991).  If WIMPs were,
like asteroids, also driven into the Sun, they would be captured by the
Sun and hence would be unavailable for capture by the Earth.

	Second, Damour \& Krauss (1998, 1999) have shown that WIMPs captured
by the outer layers of the Sun into highly eccentric orbits could evolve
into non-Sun-crossing orbits before they again collided with solar nuclei.
If so, WIMPs on these eccentric orbits could substantially increase the
number of low-velocity WIMPs in the solar neighborhood and so dramatically
increase the capture rate in the mass range 
$60\,\gev\la M_x\la 130\,\gev$ (Bergstrom et al.\ 1999).  Below 60 GeV,
WIMP capture is dominated by the iron resonance while above 130 GeV, 
$v_{\rm cut}>v_\oplus$, the minimum speed relative to the Earth for WIMPs
on highly eccentric orbits.

	Here we assess the problem of interpreting neutrino-detection
experiments in light of these two conflicting developments.

\section{Basic Approach}

	The central experimental fact that must guide our analysis is
that to date no neutrinos have been detected from the center of either
the Earth or the Sun.  The experiments therefore place upper limits on 
WIMPs.  Thus, a conservative interpretation of these experiments requires
that one assume that only those parts of velocity space are populated as
can be justified based on very secure theoretical arguments.  This
perspective implies that we must assume that {\it all} bound WIMPs with Earth
crossing orbits are driven into the Sun within a few Myr.  This includes 
primordial WIMPs, the gravitationally diffused WIMPs of Gould (1991), and
the solar-collision WIMPs of Damour \& Krauss (1998, 1999).

	Before continuing, we wish to emphasize that it is by no means
proven that all bound WIMPs are in fact driven into the Sun.  The numerical
integrations carried out to date (Farinella et al.\ 1994; Gladman et al.\ 
1997; Migliorini et al.\ 1998) 
apply to a very special subclass of Earth-crossing
orbits, namely those of existing minor bodies.  Near-Earth asteroids
are believed to be transported from their reservoir in the asteroid belt by
means of resonances.  It therefore may not be surprising that their
continued orbital evolution is dominated by resonances.  The main population
of Earth-crossing WIMPs, which acquire their orbits by quite different,
non-resonant paths (Gould 1991; Damour \& Krauss 1998,1999), could be
virtually unaffected by the resonances that drive asteroids into the
Sun.  Our viewpoint is simply that in the absence of proof that
Earth-crossing WIMPs are not depopulated, one cannot place reliable
upper limits on WIMPs from the failure to detect the annihilation signal
that would be triggered by the capture of these Earth-crossing WIMPs.

\section{Assured WIMP Capture}

	We begin by adopting an ultra-conservative view and assume 
that all bound WIMPs acquired by the solar system are immediately 
driven into the Sun.  Then, for $v_{\rm cut}<(2^{1/2}+1)v_\oplus$, it is
straight forward to show from equation (\ref{eqn:capture1}) that
\begin{equation}
C_{\rm ultra} = {2\pi N_A \sigma_{xA} f_\oplus(0)\over \beta_-}
\int_{u=v_{\rm hole}-v_\oplus}^{v_{\rm cut}}
d u^2(v_{\rm cut}^2 - u^2)\biggl(1 - {v_{\rm hole}^2 - v_\oplus^2-u^2\over
2 u v_\oplus}\biggr).
 \label{eqn:cultra}   
\end{equation}
where
\begin{equation}
v_{\rm hole} = 2^{1/2} v_\oplus.
 \label{eqn:vhole1}   
\end{equation}
Hence, the ratio of the number of WIMPs captured under ultra-conservative
assumptions to the naive formula (\ref{eqn:cap0}) is
\begin{equation}
{C_{\rm ultra}\over C_0} = \biggl\{{1\over 2}(1-s^2) +
{v_{\rm cut}\over v_\oplus}\biggl[-t^2(1-s) + {(1+t^2)(1-s^3)\over 3}
- {1-s^5\over 5}\biggr]\biggr\}\Theta(1-s),
 \label{eqn:cratio1}   
\end{equation}
where
\begin{equation}
s = {v_{\rm hole} - v_\oplus\over v_{\rm cut}}=
0.41{v_\oplus\over v_{\rm cut}},
 \quad
t = {\sqrt{v_{\rm hole}^2 - v_\oplus^2}\over v_{\rm cut}}
= {v_\oplus\over v_{\rm cut}},
 \label{eqn:stdefs}   
\end{equation}
and $\Theta$ is a step function.  The ratio $C_{\rm ultra}/C_0$ as a function
of WIMP mass $M_x$ is shown as a solid line in Figure \ref{fig:one}.

	Equation (\ref{eqn:cultra}) is in fact too conservative.  Gould (1991)
showed that Jupiter-crossing orbits (including those that also cross the
Earth's orbit) are populated from the reservoir of Galactic WIMPs on
very short time scales.  To an adequate approximation, the Jupiter-crossing
orbits can be described as those with Earth-crossing velocities $\bf u$
constrained by $|{\bf u} + {\bf v}_\oplus|>[2(1-a_\oplus/a_J)]^{1/2}v_\oplus$,
where $a_J/a_\oplus\simeq 5.2$ is the ratio of the orbital radii of Jupiter
and the Earth.  Hence the ratio of capture rates in this conservative
(but not ultra-conservative) framework is still given by equation
(\ref{eqn:cratio1}) but with
$v_{\rm hole}^2 \rightarrow 2(1-a_\oplus/a_J)v_\oplus^2$ and consequently
\begin{equation}
s = 0.27{v_\oplus\over v_{\rm cut}},
 \quad
t = 0.78{v_\oplus\over v_{\rm cut}}.
 \label{eqn:stdefs2}   
\end{equation}
This result is shown as a bold curve in Figure \ref{fig:one}.

	In principle one should take into account the loss of coherence
in WIMP-nucleon interactions that occurs when the momentum transfer, $q$, 
becomes comparable to (or larger than) the inverse radius of the nucleus,
$R\sim 3.7\,$fm for iron.  The suppression due to loss of coherence is
$\exp (-q^2 R^2/3\hbar^2)$ (Gould 1987).  
However, for WIMPs of mass $M_x$, the maximum
momentum transfer that leads to capture is 
$q_{\rm max}=(M_{\rm Fe}M_x)^{1/2}v_{\rm cut}$.  In the high mass limit, 
$q_{\rm max}
\rightarrow 2 M_{\rm Fe} v_{\rm esc}$.  Hence, the most extreme suppression
factor is $\exp[-(2 M_{\rm Fe}v_{\rm esc}R_{\rm Fe}/\hbar)^2/3]\sim 0.997$.
We conclude that loss of coherence can be ignored.  See also Gould (1987).

\section{Discussion}

	From Figure \ref{fig:one}, we see that under the conservative
assumption that WIMPs do not populate bound orbits (unless they are
Jupiter-crossing), WIMP capture is highly suppressed for WIMP mass 
$M_x\ga 150\,\gev$ and completely eliminated for $M_x>630\,\gev$.  These
results imply that, at present, the non-detection of neutrinos coming from
the Earth's center cannot be used to place limits on WIMPs of mass
$M_x>630\,\gev$.  Moreover, for masses $75\,\gev\la M_x\la 630\,\gev$, the
limits must be softened relative to what would be obtained from equation
(\ref{eqn:cap0}).

	It is quite possible that WIMPs are not generically driven into
the Sun.  The evidence that they are comes from numerical integration
of asteroid orbits, and these latter could occupy a very special
locus in parameter space.  It will be necessary to integrate typical
WIMP orbits to find out if WIMPs survive longer than asteroids.
In particular, these integrations should focus on the highly eccentric
orbits for which Damour \& Krauss (1998,1999) predict a huge enhancement
for $M_x\la 130\,\gev$.
If typical WIMP orbits are found to survive substantially
longer than asteroid orbits,
then the limits derived from the naive calculations of Gould (1987, 1991) would
become valid.  It is even possible that the much stronger limits
derived by Bergstrom et al.\ (1999) would apply.

\bigskip 

{\bf Acknowledgements}: 
We thank A.\ Quillen for pointing out the importance of recent work on
asteroid orbits.
This work was supported in part by grant AST 97-27520 from the NSF.

\clearpage 
\begin{figure}
\caption[junk]{\label{fig:one}
Conservative capture rates for WIMPs relative to the rate based on the naive 
assumption
(Gould 1991) that the phase-space density of WIMPs bound to the Sun is
similar to that of low-velocity unbound WIMPs.  The solid curve shows the
suppression factor under the ultra-conservative assumption that all bound
WIMPs are driven into the Sun on short time scales (as is true of many
Earth-crossing asteroids).  The bold curve results from the more realistic
assumption that WIMPs on Jupiter-crossing (and Earth-crossing) orbits are
repopulated faster than they can be driven into the Sun.
}
\end{figure}

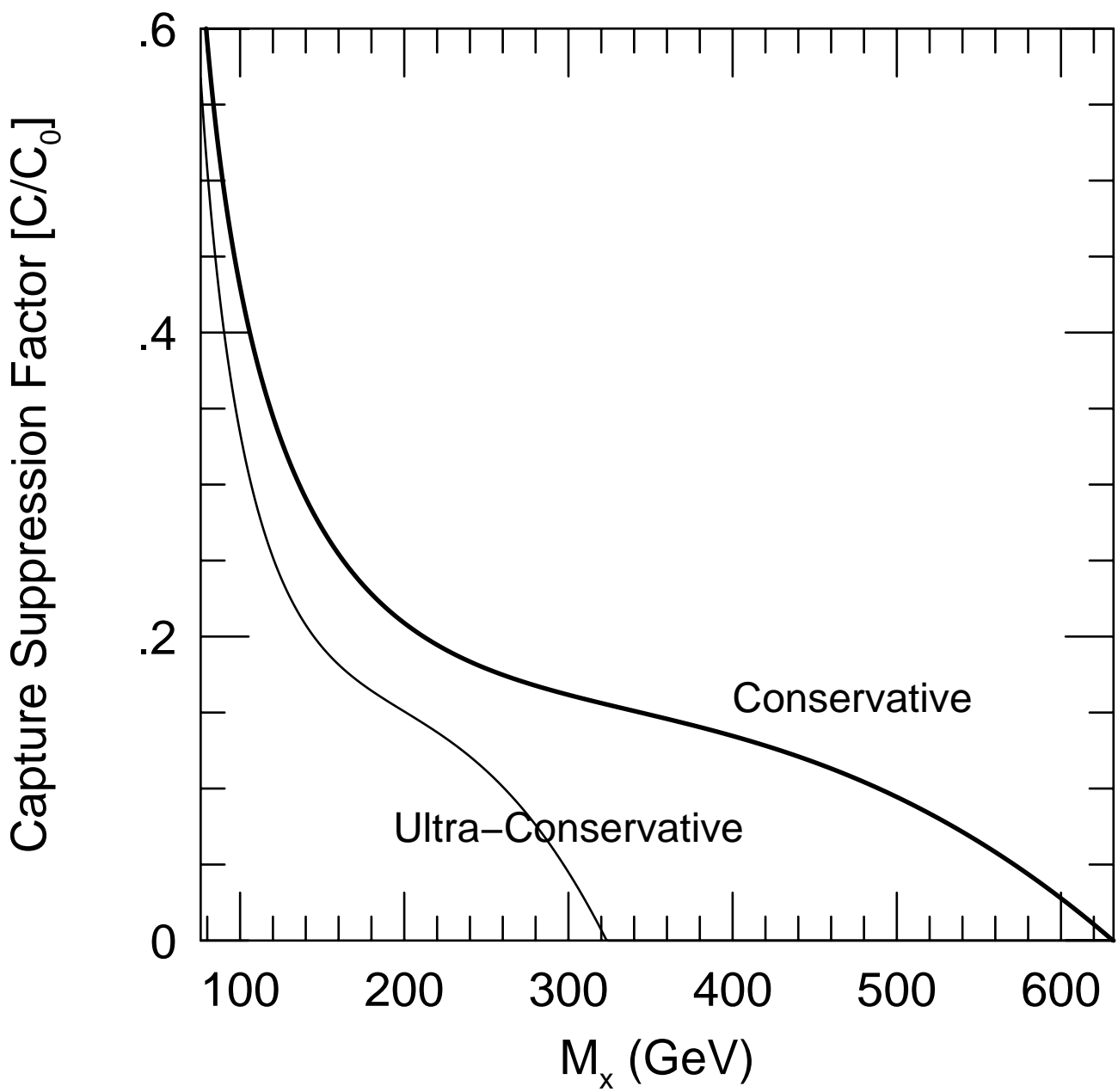
\end{document}